\documentclass[epj]{svjour}
\usepackage{graphicx}
\usepackage{amsmath,amsfonts,amssymb}
\usepackage{mathrsfs}
\usepackage{extarrows}
\usepackage{cases}
\usepackage{color}
\newcommand{\rmnum}[1]{\romannumeral #1}

\begin{document}
\title{Global and partitioned reconstructions of undirected \\
complex networks}

\author{Ming Xu\inst{1,2} \and 
Chuan-Yun Xu\inst{1} \and 
Huan Wang\inst{3} \and 
Yong-Kui Li\inst{1} \and
Jing-Bo Hu\inst{1} \and
Ke-Fei Cao\inst{1,}
\thanks{e-mail: kfcao163@163.com}%
}                     
%
%
\institute{
Center for Nonlinear Complex Systems, 
Department of Physics, 
School of Physics and Astronomy, 
Yunnan University, \\
Kunming, Yunnan 650091, P.R. China
 \and 
School of Mathematical Sciences, 
Kaili University, 
Kaili, Guizhou 556011, P.R. China
 \and 
School of Computer Science and Technology, 
Baoji University of Arts and Sciences, 
Baoji, Shaanxi 721016, P.R. China
}
%
\date{{\color{blue}Eur. Phys. J. B} (2016) {\bf 89}(3): 55\\
Received 12 December 2015 / Received in final form 18 January 2016 / Published online 2 March 2016\\
The final publication is available at Springer via 
{\color{blue}http://dx.doi.org/10.1140/epjb/e2016-60956-2}}
%
\abstract{
It is a significant challenge to predict the network topology from a small amount of dynamical observations. Different from the usual framework of the node-based reconstruction, two optimization approaches (i.e., the global and partitioned reconstructions) are proposed to infer the structure of undirected networks from dynamics. These approaches are applied to evolutionary games occurring on both homogeneous and heterogeneous networks via compressed sensing, which can more efficiently achieve higher reconstruction accuracy with relatively small amounts of data. Our approaches provide different perspectives on effectively reconstructing complex networks.
\PACS{
      {89.75.Hc}{Networks and genealogical trees}   \and
      {89.75.Fb}{Structures and organization in complex systems}   \and
      {02.50.Le}{Decision theory and game theory}   \and
      {05.45.Tp}{Time series analysis}
     } 
} 
\maketitle
\section{Introduction}

Network dynamics is the study of networks that change with time~\cite{MPB2014np}. Much evidence, from evolutionary games~\cite{PTN2006prl}, gene regulatory networks~\cite{STD2007s}, epidemic spreading and information diffusions~\cite{YWR2007pla}, transportation and communication processes~\cite{K2013pa}, etc., indicates that the topological structure of networks plays an important role in the dynamical behavior of networks. In addition, the network structure is also a basis for understanding and controlling complex networked systems~\cite{LSB2011n,YZD2013nc,XXW2015epjb}. However, often we can only obtain limited data from the dynamics of the individual units of the network, and are incapable of directly accessing the coupling strengths between the units and obtaining the underlying network topology. How to infer the interaction topology from the collective dynamics of a complex network has recently attracted extensive attention. To address this inverse problem, many methods have been proposed and they usually show robust and high performance with appropriate observations~\cite{NS2008pre,GSP2009pnas,ST2011njp,WYL2011epl,HKK2011prl,WLG2011prx,MCK2012nm,BB2013nb,SWF2014nc,TC2014jpa,HSW2015prl,LA2015pre,CLL2015pre}. In view of the issues regarding measurement cost, however, further studies to improve the prediction efficiency are necessary.

In general, two steps are followed in reconstructing the topology of sparsely connected dynamical networks: 
\linebreak[4](\rmnum{1}) recovering local structures centered at each node by optimization methods such as compressed sensing~\cite{WYL2011epl,WLG2011prx}, the lasso~\cite{HSW2015prl}, regression and Bayesian inference~\cite{MCK2012nm}; 
\linebreak[4](\rmnum{2}) assembling networks from these local structures. This node-based reconstruction (NR) approach may miss some useful information. 
When an undirected network is reconstructed by the NR, for example, the fact is not taken into account that the link from node $i$ to node $j$ should be the same as that from node $j$ to node $i$ (which can be expressed as the constraint conditions like $a_{i,j}=a_{j,i}$ in Eq.~(\ref{Gi.aij.GR}b)). Does this influence the network reconstruction? In this work, we will explore this issue. Here we propose two improved approaches to better infer the network topology under constraint conditions: the global reconstruction (GR, which considers all nodes as a whole) and the partitioned reconstruction (PR, which considers nodes in groups), respectively.

The remainder of this paper is arranged as follows. First, we recall the usual NR approach via compressed sensing. Second, we apply our reconstruction approaches (i.e., GR and PR) to two representative dynamics, including the prisoner's dilemma game (PDG) and the snowdrift game (SG) which occur on a variety of models and real complex networks. The GR and the PR of these complex networks show better results than the NR. Finally, more comparisons of these reconstruction approaches and possible extensions are discussed.

\section{Node-based reconstruction (NR)}

Compressed sensing~\cite{CRT2006tit,CRT2006cpam,D2006tit} is an effective method for signal recovery from highly incomplete information, which has broad applications in various sparse reconstruction problems. One of its main results states that a sparse vector $\mathbf{X}_{0}\in \mathbb{R}^{N}$ can be recovered from a small number of linear measurements $\mathbf{Y}=\varPhi\mathbf{X}_{0}\in \mathbb{R}^{M}$ by solving the convex program 
\begin{equation}
\label{P1}
(\text{P}_{1}) \quad \min \mathbf{\|X\|}_{1}\quad \text{subject~to}\quad \mathbf{Y}=\varPhi \mathbf{X},
\end{equation}
where $\varPhi$ is an $M\times N$ sensing matrix with $M\ll N$, and $\mathbf{\|X\|}_{1}:=\sum_{i=1}^{N}|X_{i}|$ the $\ell_{1}$ norm of $\mathbf{X}$. When all the entries of $\mathbf{X}$, $\varPhi$ and $\mathbf{Y}$ are real-valued, $(\text{P}_{1})$ can eventually be solved by the linear programming. In fact, many real networks are sparse, which makes it possible to determine the structure of networks based on the small amount of data from dynamics via compressed sensing~\cite{WLG2011prx,SWF2014nc}. For comparison, some classic examples of the NR via compressed sensing will be introduced in this section~\cite{WLG2011prx}.

In game theory, agents (or players) use different strategies in order to gain the maximum payoff. Here, the strategies can be divided into two types: cooperation ($C$) and defection ($D$), denoted by $\mathbf{S}(C)=(1,0)^{T}$ and $\mathbf{S}(D)=(0,1)^{T}$ respectively, where $T$ stands for the transpose operation. The payoffs of the two agents in a game are determined by their strategies and the payoff matrix of the specific game. For the PDG~\cite{NM1992n} and the SG~\cite{HD2004n}, the payoff matrices can be presented as: 
\begin{equation}
\label{P.matrices}
\begin{array}
[c]{ccc}%
P_\text{PDG}=\left(
\begin{array}
[c]{cc}%
1 & 0\\
b & 0
\end{array}
\right) & \text{and} & 
P_\text{SG}=\left(
\begin{array}
[c]{cc}%
1 & 1-r\\
1+r & 0
\end{array}
\right),
\end{array}
\end{equation}
respectively, where $b$ ($1<b<2$) and $r$ ($0<r<1$) are parameters characterizing the temptation to defect. At each time step, all agents play the game with their neighbors. For agent $i$, the payoff is: 
\begin{equation}
\label{gi}
g_{i}=\sum\limits_{j\in\Gamma_{i}} \mathbf{S}_{i}^{T}P\mathbf{S}_{j},
\end{equation}
where $\mathbf{S}_{i}$ and $\mathbf{S}_{j}$ denote the strategies of agents $i$ and $j$ respectively, and $\Gamma_{i}$ is the neighbor-connection set of $i$. According to the Fermi rule, after each round, agent $i$ randomly chooses a neighbor $j$, and switches to the strategy of $j$ with the probability~\cite{ST1998pre} 
\begin{equation}
\label{w}
w_{i\leftarrow j}=\frac{1}{1+\exp[(g_{i}-g_{j})/\kappa]},
\end{equation}
where the noise parameter $\kappa$ characterizes the level of uncertainty by strategy adoptions~\cite{SF2007pr}. The interactions among agents in the network can be characterized by an $N\times N$ adjacency matrix $A$ with entries $a_{i,j}=1$ if agents $i$ and $j$ are connected, and $a_{i,j}=0$ otherwise. Here, $a_{i,i}$ is always treated as $0$. The payoff of agent $i$, at the $t$th round, can be expressed as: 
\begin{equation}
\label{gi.t}
g_{i}(t)=\sum\limits_{j=1}^{N} a_{j,i}\mathbf{S}_{i}^{T}(t)P\mathbf{S}_{j}(t).
\end{equation}
When $t=1,2,\ldots,M$, we get $M$ equations from equation~(\ref{gi.t}), which can be grouped as: 
\begin{equation}
\label{c.Gi}
\mathbf{G}_{i}=\varPhi_{i} \mathbf{A}_{i},
\end{equation}
where $\mathbf{G}_{i}=(g_{i}(1),g_{i}(2),\ldots,g_{i}(M))^{T}$, $\varPhi_{i}$ is an $M\times N$ matrix with entries $(\varPhi_{i})_{t,j}=\mathbf{S}_{i}^T(t)P\mathbf{S}_{j}(t)$, and $\mathbf{A}_{i}$ is the $i$th column vector of $A$. Since payoffs and strategies in equation~(\ref{gi.t}) are observable, $\mathbf{G}_{i}$ and $\varPhi_{i}$ are known. $\mathbf{A}_{i}$ is sparse due to the natural sparsity of complex networks, which would ensure a conversion from the local structure reconstruction into a sparse signal reconstruction. Thus we can predict $\mathbf{A}_{i}$ from equation~(\ref{c.Gi}) by compressed sensing. Similarly, other columns of $A$ can also be inferred, hence the reconstructed adjacency matrix $A^\text{NR}$ is obtained. The entries in the predicted $A^\text{NR}$ may not be exactly $0$ or $1$. If $a_{i,j}^\text{NR}$, the predicted value of an entry $a_{i,j}$, is close to $1$, then a link from $i$ to $j$ is predicted to be existent; if $a_{i,j}^\text{NR}$ is close to $0$, then a null (nonexistent) link is predicted. Here, a threshold $0.5$ is taken to predict the network structure: the entries in $A^\text{NR}$ less than $0.5$ ($a_{i,j}^\text{NR}<0.5$) are considered as null links, and the rest ($a_{i,j}^\text{NR}\geqslant 0.5$) are regarded as the existence of the corresponding links.

To evaluate the performance of predictions of network structures, we adopt the success rates of existent links (SREL) and nonexistent links (SRNL)~\cite{WLG2011prx,SWF2014nc}. SREL \linebreak[4](SRNL) is defined as the ratio of the number of successfully predicted existent (nonexistent) links to the total number of existent (nonexistent) links. The PDG and the SG on networks discussed above are two typical models of games. In fact, the evolutionary games on complex networks can exhibit very rich dynamical behavior~\cite{SVS2005pre,PSS2008pre,PS2010bs,WSP2012sr}. The actual topology, as well as payoffs and noise, could have a significant role on the dynamics. In particular, the presence of overlapping triangles on the two-dimensional lattice structures~\cite{SVS2005pre} or restricted connections among some players~\cite{PSS2008pre} is essential. In our study here, for the dynamical processes of PDG and SG on the three types of network topologies (including Erd\H{o}s-R\'{e}nyi (ER) random~\cite{ER1959pmd}, Watts-Strogatz (WS) small-world~\cite{WS1998n}, and Barab\'{a}si-Albert (BA) scale-free~\cite{BA1999s} networks), we record strategies and payoffs of agents at different times to apply the above reconstruction method (NR) with respect to different amounts of data (which can be explained as $\text{Data}=M/N$, where $M$ is the number of accessible time instances in the time series, $N$ the size (total number of nodes) of networks to be recovered).

\begin{figure*}[!]
\centering
\includegraphics[width=7in]{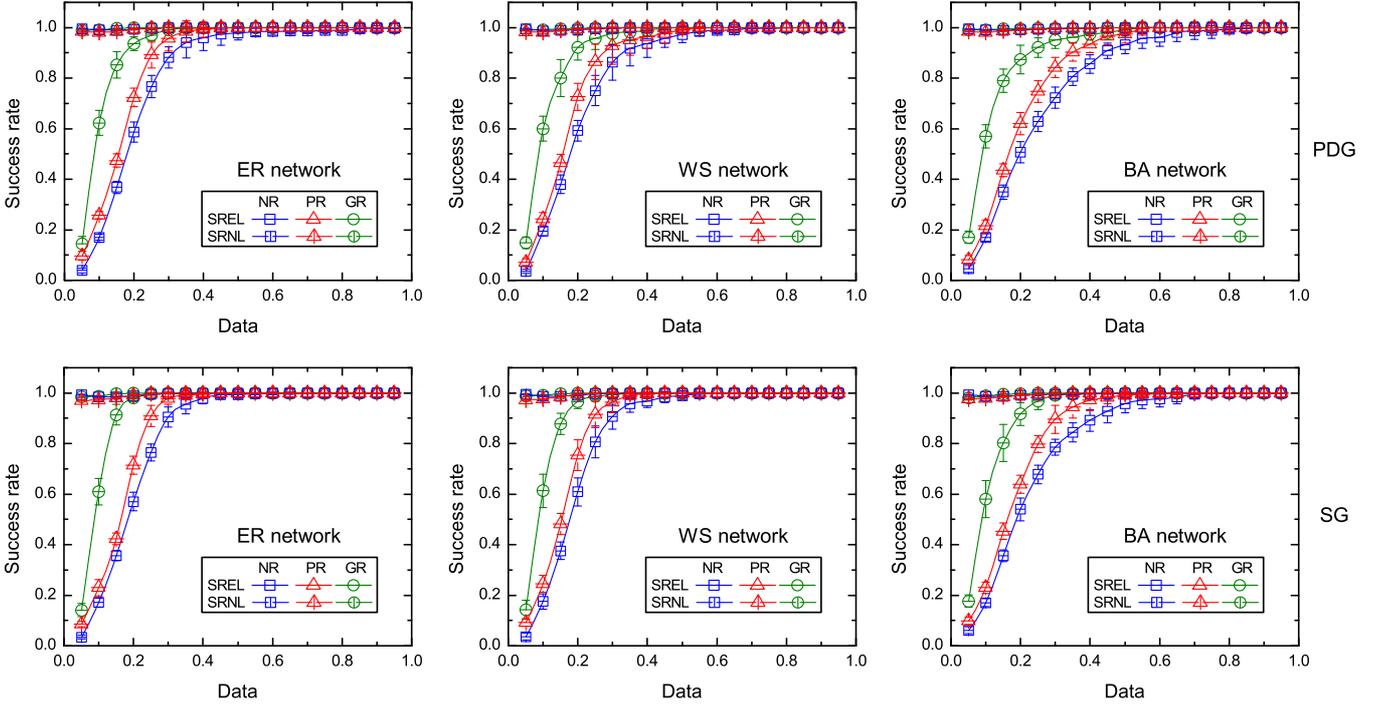}
\caption{\label{fig.1}
Comparisons of success rates of inferring three types of networks (including ER random, WS small-world and BA scale-free networks) by the NR, PR and GR, with PDG and SG dynamics, respectively. The network size $N$ is $100$. Each data point is obtained by averaging over $10$ network realizations. The error bars denote the standard deviations. The payoff parameters for the PDG and the SG are $b=1.2$ and $r=0.7$, respectively. The average node degrees of all used networks are fixed to $6$ and the noise parameter $\kappa=0.1$. We can see that with the same and relatively small amounts of data, the GR achieves the highest success rates, and the PR also achieves higher success rates than the NR. The differences of success rates of the NR, PR and GR become not obvious when the amounts of data are large enough.
}
\end{figure*}

The examples discussed above show a typical procedure of the NR. According to equation~(\ref{c.Gi}), the neighbors of node $i$ can be discovered, and the local structures of other nodes in the network can be obtained similarly. Thus, the topology of the whole network can be determined eventually, which can achieve good predictions as shown in Figure~\ref{fig.1}.

\section{Global reconstruction (GR)}

Some puzzles still remain because the constraint of $a_{i,j}=a_{j,i}$ in adjacency matrix $A$ of an undirected complex network cannot be reflected in the NR procedure. So, how will such information be taken into account? For this, we first propose a GR approach which makes full use of the known information and can be used to mine the interactions between nodes. We will test its performance using the above evolutionary games.

First, taking all nodes as a whole, we collect all the possible constraints from the dynamics, which can be expressed as: 
\begin{subequations}
\label{Gi.aij.GR}
\begin{numcases}{}
\mathbf{G}_{i}=\varPhi_{i} \mathbf{A}_{i}\quad (i=1,2,\ldots,N),\\
a_{i,j}=a_{j,i}\quad (i,j=1,2,\ldots,N),
\end{numcases}
\end{subequations}
where equations (\ref{Gi.aij.GR}a) come from equation~(\ref{c.Gi}), and constraints (\ref{Gi.aij.GR}b) from the characteristics of the evolutionary games. Second, we incorporate the reconstruction problem into the framework of compressed sensing by transforming equation~(\ref{Gi.aij.GR}) into the form like $\mathbf{Y}=\varPhi \mathbf{X}$ in equation~(\ref{P1}). It would be helpful to rewrite equations (\ref{Gi.aij.GR}a) as: 
\begin{equation}
\label{G.Phi.A}
\left(\begin{array}{c}
\mathbf{G}_{1}\\
\mathbf{G}_{2}\\
\vdots\\
\mathbf{G}_{N}
\end{array}\right)
=
\left(\begin{array}{cccc}
\varPhi_{1}&&&\\
&\varPhi_{2}&&\\
&&\ddots&\\
&&&\varPhi_{N}
\end{array}\right)
\left(\begin{array}{c}
\mathbf{A}_{1}\\
\mathbf{A}_{2}\\
\vdots\\
\mathbf{A}_{N}
\end{array}\right)
\end{equation}
or
 \begin{equation}
\label{tilde.G.1}
\widetilde{\mathbf{G}}=\widetilde{\varPhi} \widetilde{\mathbf{A}}.
\end{equation}
Let us denote the column vectors of $\widetilde{\varPhi}$, in sequence, as $\widetilde{\mathbf{\Phi}}_{1},\widetilde{\mathbf{\Phi}}_{2},\ldots,\widetilde{\mathbf{\Phi}}_{N^{2}}$. Then, equation~(\ref{tilde.G.1}) can be transformed as: 
\begin{equation}
\label{tilde.G.2}
\widetilde{\mathbf{G}}=\sum \limits_{i,j=1}^{N} \widetilde{\mathbf{\Phi}}_{j+(i-1)N}a_{j,i}.
\end{equation}
Considering $a_{i,j}=a_{j,i}$ in equation~(\ref{Gi.aij.GR}b), we can simplify equation~(\ref{tilde.G.2}), by combining like terms, as: 
\begin{equation}
\label{tilde.G.3}
\widetilde{\mathbf{G}}=\sum \limits_{1\leqslant i<j\leqslant N}(\widetilde{\mathbf{\Phi}}_{j+(i-1)N}+\widetilde{\mathbf{\Phi}}_{i+(j-1)N}) a_{j,i},
\end{equation}
which can be recast as: 
\begin{equation}
\label{tilde.G.4}
\widetilde{\mathbf{G}}=\widetilde{\varPhi}^{-}\widetilde{\mathbf{A}}^{-},
\end{equation}
where $\widetilde{\mathbf{G}}$ and $\widetilde{\varPhi}^{-}$ (whose entries come from $\widetilde{\varPhi}$) are measurable, and the vector 
\begin{equation}
\label{tilde.A.m}
\begin{array}{ccccc}
\widetilde{\mathbf{A}}^{-}=&(a_{2,1},&{\ \ }a_{3,1},&{\ }\ldots,&a_{N,1},\\
&&{\ \ }a_{3,2},&{\ }\ldots,&a_{N,2},\\
&&&{\ }\ldots,&\ldots,\\
&&&&a_{N,N-1})^{T}
\end{array}
\end{equation}
is sparse. Finally, $\widetilde{\mathbf{A}}^{-}$ can be predicted from equation~(\ref{tilde.G.4}) by compressed sensing, hence $\widetilde{\mathbf{A}}$ and adjacency matrix $A$ can be consequently predicted.

\begin{table*}
\caption{\label{tab.1}
Minimum data for achieving at least $0.95$ SREL and SRNL simultaneously for PDG and SG dynamics on different networks by the GR, PR and NR. Here, $N$ is the network size, and $\langle k\rangle $ the average degree. The results of the PR are obtained by dividing all nodes of every network into $5$ groups with equal (or approximately equal) number of nodes. Each of $\text{Data}$ in the last two columns is an average over $10$ independent realizations. Simulations are also performed on two real networks, i.e., the social network of dolphins~\cite{LSB2003bes} and the network reflecting the schedule of college (American) football games~\cite{GN2002pnas}.
}
\begin{center}
\begin{tabular}{ccccccccc}
\hline
          &&      &&      &&      PDG      &&      SG\\
Network   && $N$  && $\langle k\rangle $ && (GR / PR / NR) && (GR / PR / NR)\\
\hline
ER        &&  60  &&  4   &&  0.25 / 0.43 / 0.50  &&  0.20 / 0.38 / 0.44\\
          &&  60  &&  6   &&  0.24 / 0.42 / 0.48  &&  0.24 / 0.39 / 0.45\\
          && 100  &&  6   &&  0.21 / 0.31 / 0.37  &&  0.18 / 0.29 / 0.35\\
          && 100  && 12   &&  0.22 / 0.42 / 0.48  &&  0.21 / 0.40 / 0.46\\
WS        &&  60  &&  4   &&  0.31 / 0.48 / 0.55  &&  0.24 / 0.38 / 0.44\\
          &&  60  &&  6   &&  0.30 / 0.46 / 0.52  &&  0.22 / 0.44 / 0.51\\
          && 100  &&  6   &&  0.24 / 0.36 / 0.42  &&  0.19 / 0.30 / 0.35\\
          && 100  && 12   &&  0.24 / 0.37 / 0.43  &&  0.21 / 0.37 / 0.43\\
BA        &&  60  &&  4   &&  0.28 / 0.53 / 0.60  &&  0.23 / 0.47 / 0.54\\
          &&  60  &&  6   &&  0.30 / 0.49 / 0.56  &&  0.25 / 0.48 / 0.55\\
          && 100  &&  6   &&  0.29 / 0.45 / 0.53  &&  0.24 / 0.42 / 0.49\\
          && 100  && 12   &&  0.30 / 0.48 / 0.56  &&  0.26 / 0.46 / 0.54\\
Dolphins  &&  62  &&  5.1 &&  0.32 / 0.56 / 0.65  &&  0.24 / 0.50 / 0.58\\
Football  && 115  && 10.7 &&  0.29 / 0.44 / 0.50  &&  0.20 / 0.32 / 0.37\\
\hline
\end{tabular}
\end{center}
\end{table*}

We will validate the GR approach by extensive numerical computations for typical models and real complex networks. From Figure~\ref{fig.1} and Table~\ref{tab.1} we can see that, even if the available information about each agent's strategy and payoff is very limited, our approach is quite efficient in predicting network links. When achieving at least $0.95$ SREL and SRNL simultaneously in the reconstructions, the observation data required for the GR are only roughly half that for the NR. For heterogeneous networks such as BA scale-free networks, the NR is difficult to achieve high prediction accuracy because some high-degree hub nodes violate the local sparsity. Whereas the GR works well since the global sparsity is maintained.

For many complex networks, although the GR greatly enhances our ability to infer the network topology, it may result in the problem of high dimensionality. For the NR, we know that $\varPhi_{i}$ is an $M\times N$ matrix from equation~(\ref{c.Gi}); while for the GR, $\widetilde{\varPhi}^{-}$ is an $MN\times (N(N-1)/2)$ matrix from equation~(\ref{tilde.G.4}). So, the GR requires much more computer memory which mainly depends on the network size $N$.

\section{Partitioned reconstruction (PR)}

Compared with the NR, the GR greatly improves the accuracy of the prediction, but also increases the size of the problem. Here, we propose a more flexible reconstruction scheme (i.e., PR), which is based on the information 
\begin{subequations}
\label{Gi.aij.PR}
\begin{numcases}{}
\mathbf{G}_{i}=\varPhi_{i} \mathbf{A}_{i}\quad (i\in\alpha\subseteq\{1,2,\ldots,N\}),\\
a_{i,j}=a_{j,i}\quad (i,j\in\alpha\subseteq\{1,2,\ldots,N\}).
\end{numcases}
\end{subequations}
This means that the nodes can be treated in groups, so the PR is in fact a group-based reconstruction. Usually, the more elements in $\alpha$, the better the reconstruction. So, if the computing environment is allowed, we should implement the PR with as many as possible elements in $\alpha$. Especially, when $\alpha=\{i\}$, equation~(\ref{Gi.aij.PR}) is equal to equation~(\ref{c.Gi}); and when $\alpha=\{1,2,\ldots,N\}$, equation~(\ref{Gi.aij.PR}) is equal to equation~(\ref{Gi.aij.GR}). Thus, the NR and the GR can be regarded as the smallest- and the largest-scale PR, respectively.

Taking the above collective dynamics for example, we can specify the PR approach in three steps as follows:

(\rmnum{1}) According to the actual specific problems, algorithm design, computer memory, etc., the maximum capacity allowed for each group is determined, thus all nodes in the network can be properly grouped. Here, we consider a network with $N=100$ nodes. Suppose that $20$ is the maximum number of nodes for each group, then all nodes of the original network can be randomly divided into $5$ groups with the same size, and they can be denoted as: $\alpha_{1}=\{1,2,\ldots,20\},\alpha_{2}=\{21,22,\ldots,40\},\ldots,\alpha_{5}=\{81,82,\ldots,100\}$.

(\rmnum{2}) Based on reconstructing each group as a whole, the structure of the network can be inferred. For group $\alpha_{1}$, we obviously have 
\begin{subequations}
\label{Gi.aij.alpha1}
\begin{numcases}{}
\mathbf{G}_{i}=\varPhi_{i} \mathbf{A}_{i}\quad (i\in\alpha_{1}=\{1,2,\ldots,20\}),\\
a_{i,j}=a_{j,i}\quad (i,j\in\alpha_{1}=\{1,2,\ldots,20\}).
\end{numcases}
\end{subequations}
Equations (\ref{Gi.aij.alpha1}a) can be rewritten as: 
\begin{equation}
\label{G.Phi.A.alpha1}
\left(\begin{array}{c}
\mathbf{G}_{1}\\
\mathbf{G}_{2}\\
\vdots\\
\mathbf{G}_{20}
\end{array}\right)
=
\left(\begin{array}{cccc}
\varPhi_{1}&&&\\
&\varPhi_{2}&&\\
&&\ddots&\\
&&&\varPhi_{20}
\end{array}\right)
\left(\begin{array}{c}
\mathbf{A}_{1}\\
\mathbf{A}_{2}\\
\vdots\\
\mathbf{A}_{20}
\end{array}\right)
\end{equation}
or 
 \begin{equation}
\label{tilde.G.1.alpha1}
\bar{\mathbf{G}}=\bar{\varPhi}\bar{\mathbf{A}}.
\end{equation}
Denote the column vectors of $\bar{\varPhi}$, in sequence, as $\bar{\mathbf{\Phi}}_{1},\bar{\mathbf{\Phi}}_{2},$ $\ldots,\bar{\mathbf{\Phi}}_{2000}$. Then, equation~(\ref{tilde.G.1.alpha1}) can be transformed as: 
\begin{equation}
\label{tilde.G.2.alpha1}
\bar{\mathbf{G}}=\sum \limits_{1\leqslant i\leqslant 20, 1\leqslant j\leqslant 100} \bar{\mathbf{\Phi}}_{j+100(i-1)}a_{j,i}.
\end{equation}
Considering $a_{i,j}=a_{j,i}$ in equation~(\ref{Gi.aij.alpha1}b), we can simplify equation~(\ref{tilde.G.2.alpha1}), by combining like terms, as: 
\begin{equation}
\label{tilde.G.3.alpha1}
\begin{array}{ll}
\bar{\mathbf{G}}=&\sum \limits_{1\leqslant i<j\leqslant 20}(\bar{\mathbf{\Phi}}_{j+100(i-1)}+\bar{\mathbf{\Phi}}_{i+100(j-1)}) a_{j,i}\\
&+\sum \limits_{1\leqslant i\leqslant 20, 21\leqslant j\leqslant 100} \bar{\mathbf{\Phi}}_{j+100(i-1)}a_{j,i},
\end{array}
\end{equation}
which can be recast as: 
\begin{equation}
\label{tilde.G.4.alpha1}
\bar{\mathbf{G}}=\bar{\varPhi}^{-}\bar{\mathbf{A}}^{-},
\end{equation}
where $\bar{\mathbf{G}}$ and $\bar{\varPhi}^{-}$ are measurable, and the vector 
\begin{equation}
\label{tilde.A.m.alpha1}
\begin{array}{ccccccc}
\bar{\mathbf{A}}^{-}=&(a_{2,1},&{\ }a_{3,1},&{\ }\ldots,&\ldots,&\ldots,&a_{100,1},\\
&&{\ }a_{3,2},&{\ }\ldots,&\ldots,&\ldots,&a_{100,2},\\
&&&{\ }\ldots,&\ldots,&\ldots,&\ldots,\\
&&&&a_{21,20},&\ldots,&a_{100,20})^{T}
\end{array}
\end{equation}
is sparse. Therefore, $\bar{\mathbf{A}}^{-}$ can be predicted from equation (\ref{tilde.G.4.alpha1}) by compressed sensing, and $\bar{\mathbf{A}}$ can thus be predicted. Namely, we can get predicted values of entries for \linebreak[4]$\mathbf{A}_{1},\mathbf{A}_{2},\ldots$, $\mathbf{A}_{20}$ by the transformations like the GR. Note that $\mathbf{A}_{i}$ ($i=1,2,\ldots,20$) is the $i$th column of adjacency matrix $A$, which stands for all links of node $i$. For other groups $\alpha_{2}$, $\alpha_{3}$, $\alpha_{4}$ and $\alpha_{5}$, the remaining columns (i.e., $\mathbf{A}_{21},\mathbf{A}_{22},\ldots,\mathbf{A}_{100}$) of adjacency matrix $A$ can also be inferred similarly. So, the structure of the network is inferred, and the predicted adjacency matrix is denoted as $A^\text{PR1}$. 

(\rmnum{3}) For $A^\text{PR1}$, some reconstruction results are regarded as violating the constraint conditions if the values of \linebreak[4]the entries $a^\text{PR1}_{i,j}$ and $a^\text{PR1}_{j,i}$ in $A^\text{PR1}$ are not close enough. To assess the violation, here we introduce a cumulative deviation index for node $i$, which is defined as $d_{c}(i)=\sum_{j=1}^{N}|a^\text{PR1}_{i,j}-a^\text{PR1}_{j,i}|$. In particular, $d_{c}(i)=0$ implies that node $i$ does not violate the constraint condition $a_{i,j}=a_{j,i}$. To pursue better prediction, the top $20$ nodes with high $d_{c}$ will be picked out from the network. They, as a whole, can be reconstructed, and the re-reconstruction data can update the results in step (\rmnum{2}). Thus, we can get a newly inferred adjacency matrix denoted as $A^\text{PR2}$ which can better and more reasonably predict the network structure. Similarly, by analyzing the cumulative deviation index for $A^\text{PR2}$, one can further obtain $A^\text{PR3}$, but the improvement of the prediction is not obvious. In our numerical simulations, we can usually obtain satisfactory prediction results using only up to $A^\text{PR2}$.

\begin{figure}
\centering
\includegraphics[width=3.1in,trim=15 19 23 23,clip]{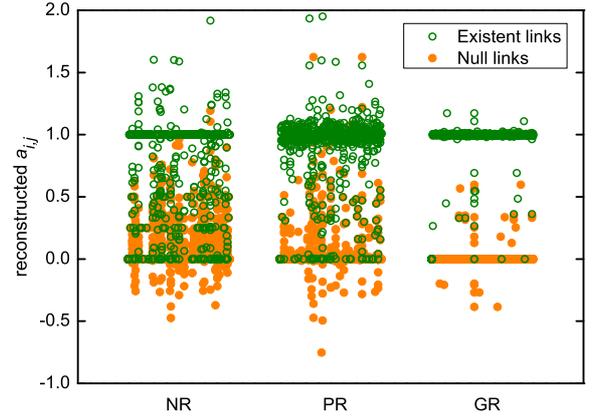}
\caption{\label{fig.2}
Reconstructed values of entries $a_{i,j}$ in adjacency matrix $A$ for the PDG on ER random networks by the NR, PR and GR, respectively. Here, $\text{Data}=0.25$, $N=100$ and $\langle k\rangle =6$. The results are obtained from two independent realizations.
}
\end{figure}

From Figures~\ref{fig.1} and \ref{fig.2}, we can see that the prediction accuracy of the PR is lower than that of the GR, but significantly higher than that of the NR. Figure~\ref{fig.2} shows the reconstruction results from the PDG on ER random networks. For $\text{Data}=0.25$, links are difficult to identify by the NR or the PR because of the mixture of reconstructed entries in $A$; whereas the GR, resulting in a clear separation between actual links and null links, can ensure nearly perfect reconstruction.

More numerical simulations illustrate that the grouping patterns in the PR will affect the success rate of reconstruction. To achieve higher prediction accuracy, it is worthy of further studies to improve the PR; here, one of the interesting questions is how to divide the nodes into groups when the maximum capacity of each group is certain.

\section{Discussion and conclusions}

To accurately and effectively infer the topological structure of complex networks, we propose the global and partitioned reconstructions (GR and PR). Their performance has been systematically analyzed and compared with the usual node-based reconstruction (NR). All these three approaches have their own advantages and disadvantages, so none of them would ensure universally the best reconstructions for various networks. Although the NR can rapidly decompose the difficult task of inferring network structure even for large-scale networks, it only focuses on the local information discovery, which leads to low success rates especially when observations are relatively rare. Besides, the performance of the NR becomes poor when the network size $N$ is small or hub nodes exist. For a network with some information that cannot be reflected by the NR, the GR may become the most effective, but this approach will encounter the curse of dimensionality when $N$ is large enough. Moreover, the PR with high flexibility offers a practical approach to deal with wide-ranging problems of network reconstructions. Although a grouping method we provided for the PR gets better results than the NR, the underlying mechanisms of the PR are not yet very clear because the grouping is very flexible and complex.

Our GR and PR approaches, to some extent, are generally effective and applicable to the reconstruction of complex networks, no matter what optimization method (compressed sensing, the lasso or other method) is used. Collective dynamics are not confined to the PDG and the SG, and the constraint conditions are not confined only to an a priori known symmetry ($a_{i,j}=a_{j,i}$) of adjacency matrices. Recently, Ma et al. proposed the conflict-based method (CBM) for efficient reconstruction of undirected heterogeneous networks with hubs~\cite{MHS2015po}, whereas our GR and PR approaches can be applied to both homogeneous and heterogeneous networks. In this paper, although we only show that the GR and PR approaches can be realized for undirected networks, one could also implement these approaches similarly for other networks if there are some other prior known characteristic properties as constraints. For example, when some links in a network are measurable in advance (say, some entries in adjacency matrix to be $1$), one could realize the PR by replacing equation~(\ref{Gi.aij.PR}b) by the known information ($a_{i,j}=1$ for the corresponding $\{i,j\}$). It would be interesting to extend the GR and PR approaches to other types of constraints (by replacing corresponding (\ref{Gi.aij.GR}b) and (\ref{Gi.aij.PR}b) respectively), such as knowing the general degree distribution in advance, or the anticipated levels of link density/clustering. Besides the accumulated payoff of agents considered in this paper, if one considers a more realistic degree-normalized payoff in the game, then the evolutionary outcome can change significantly~\cite{SPD2008pa}; in this case, how to reconstruct networks would be worth investigating further. We believe that deep data mining in dynamics would throw new light on how to achieve higher prediction accuracy from limited observations.

\section*{Author contribution statement}

MX and KFC conceived and designed the research. MX and CYX performed the analytical and numerical calculations. All authors contributed to analyses, discussions, and interpretations of the results. MX and KFC wrote the manuscript.

\bigskip

\noindent\small{This work was supported by the National Natural Science Foundation of China (NSFC) (Grant No. 11365023). MX acknowledges support from the Joint Fund of Department of Science and Technology of Guizhou Province / Bureau of Science and Technology of Qiandongnan Prefecture / Kaili University (Grant No. LH-2014-7231), the Science and Technology Talent Support Program of Department of Education of Guizhou Province (Grant No. KY-2015-505), and the Excellent Teacher Education and Training Program of Guizhou Province (Mathematics and Applied Mathematics). HW acknowledges support from the Project of the Science and Technology Research and Development Program of Baoji City (Grant No. 15RKX-1-5-14) and the Key Project of Baoji University of Arts and Sciences (Grant No. ZK14035). The authors would like to thank the anonymous referees and Dr. Xu-Sheng Zhang for their helpful comments and constructive suggestions on an earlier version of this paper.
}

%
%


\end{document}